\documentclass[12pt,preprint]{aastex}
\usepackage{emulateapj5}
\usepackage{epsf}

\newcommand{\kband}{$K'$-band}
\newcommand{\hst}{\textit{HST}}
\newcommand{\lefe}{Le F\`{e}vre}
\newcommand{\etal}{et al.}

\newcommand{\msun}{M$_{\odot}$}

\newenvironment{inlinefigure}{
\def\@captype{figure}
\noindent\begin{minipage}{0.999\linewidth}\begin{center}}
{\end{center}\end{minipage}\smallskip}

\slugcomment{To Be Published in ApJ Letters}

\shorttitle{A Slow Merger History of Field Galaxies}
\shortauthors{Bundy et al.}

\begin{document}


\title{A Slow Merger History of Field Galaxies Since $z \sim 1$\altaffilmark{1}}

\author{Kevin Bundy \altaffilmark{2}, Masataka Fukugita
\altaffilmark{3}, Richard S. Ellis \altaffilmark{2}, Tadayuki
Kodama \altaffilmark{4}, Christopher J. Conselice \altaffilmark{2}}

\smallskip

\email{kbundy@astro.caltech.edu, fukugita@sdss1.icrr.u-tokyo.ac.jp,
  rse@atro.caltech.edu, kodama@th.nao.ac.jp, cc@astro.caltech.edu}
\altaffiltext{1}{Based on data acquired at the Subaru Telescope, which is
  operated by the National Astronomical Observatory of Japan.}
\altaffiltext{2}{105--24 Caltech, 1201 E. California Blvd., Pasadena, CA 91125}
\altaffiltext{3}{Institute for Cosmic Ray Research, University of Tokyo,
  Kashiwa 277 8582, Japan}
\altaffiltext{4}{National Astronomical Observatory of Japan, Mitaka,
Tokyo 181--8588, Japan}

\smallskip

\begin{abstract}

{Using deep infrared observations conducted with the CISCO imager on the
Subaru Telescope, we investigate the field-corrected pair fraction and
the implied merger rate of galaxies in redshift survey fields with
Hubble Space Telescope imaging.  In the redshift interval, $0.5 < z <
1.5$, the fraction of infrared-selected pairs increases only modestly
with redshift to 7\% $\pm$ 6\% at $z\sim 1$.  This is nearly a factor
of three less than the fraction, 22\% $\pm$ 8\%, determined using the
same technique on {\it HST} optical images and as measured in a previous
similar study.  Tests support the hypothesis that optical pair fractions
at $z \sim 1$ are inflated by bright star-forming regions that are
unlikely to be representative of the underlying mass distribution. By
determining stellar masses for the companions, we estimate the mass
accretion rate associated with merging galaxies. At $z \sim 1$, we
estimate this to be $2\times 10^{9 \pm 0.2}$ \msun~galaxy$^{-1}$
Gyr$^{-1}$. Although uncertainties remain, our results suggest that the
growth of galaxies via the accretion of pre-existing fragments remains
as significant a phenomenon in the redshift range studied as that
estimated from ongoing star formation in independent surveys.}

\end{abstract}


\section{Introduction}

The hierarchical growth of dark matter halos is thought to govern the
assembly history and morphological evolution of galaxies.  Nearby
examples of interacting and merging galaxies are well known, and many
attempts to survey the merging and mass accretion rate at various
redshifts have been made by several groups \citep{burk94, carl94,
yee95, pat97, lef00, pat00, pat02, con03}. Strong evolution of
the global merger rate was used to explain the observed faint galaxy
excess \citep{broad92}, the evolution of the luminosity function
\citep{lilly95, ellis96}, and that of galaxy morphologies \citep{gia96,
brinch98}. Evolution of the merger rate can also be used to place
constraints on structure formation \citep{baugh96, kauf96}.

\citet{lef00} used Hubble Space Telescope ({\it HST}) F814W images of
redshift survey fields to measure the pair fraction to $z \sim 1$. They
found an increase in the field-corrected pair fraction to 20\% at $z
\sim 0.75$-1.  However, as \lefe~\etal~discuss, various biases affect this
result.  For example, in the rest-frame blue, bright star-forming
regions, possibly triggered by interactions, might inflate the
significance of pair statistics and give a false indication of the mass
assembly rate.

Infrared observations are less biased by star formation, and serve as a
better tracer of the underlying stellar mass in galaxies
\citep{broad92}.  \citet{dick03} employed this in their investigation of
the global stellar mass density for $z < 3$.  They find that 50 to 70\%
of the present-day stellar mass was in place by $z \sim 1$.  A second
line of evidence, the decline from $z \sim 1$ to 0 in the global star
formation rate (e.g., Lilly et al. 1996), provides further support for
the contention that galaxy growth was not yet complete at $z \sim 1$.
Though a chronological picture of stellar mass assembly is emerging, the
processes driving it are not understood.  Is star-formation and stellar
mass assembly induced mainly through the gradual accretion of gas
converted quiescently into stars, or does assembly occur through
merging, potentially accompanied by tidally-induced star formation?
Characterizing the continued growth of galaxies, and specifically the
contribution from galaxy mergers since $z \sim 1$, is the major goal of
this work.

\section{Observations}

In addition to high resolution infrared imaging, this study relies on
optical HST data to facilitate the comparison between infrared and
optical pair statistics and constrain stellar $M/L_K$ ratios used to
infer the stellar mass accretion rate.  We therefore selected fields for
our \kband~imaging campaign that contain a combination of
statistically-complete redshift surveys and archival \hst~F814W
imaging. Target galaxies of known redshift were selected from the
RA=10hr field of the CFRS \citep{lilly95}, which spans the apparent
magnitude range $17.5 < I_{AB} < 22.5$, and the Groth Strip area,
surveyed by the Deep Extragalactic Evolutionary Probe (DEEP: Koo 1995) and
selected according to $(R + I)/2 < 23$ (Koo 2000, private communication).

Archival \hst~images of the Groth Strip, retrieved from two programs
(GTO 5090, PI: Groth; GTO 5109, PI: Westphal), reach $I \approx 24$
\citep{gro94}, a depth sufficient for us.  The deeper \hst~images of
the CFRS fields ($I < 24.5$) are described by \citet{brinch98}. 

\kband~observations were performed using the CISCO imager \citep{moto02}
on the Subaru telescope during two campaigns in 2002 April and 2003
April. The camera has a field of 108 arcsec on a side with the pixel
size of 0.11 arcsec, and is thus fairly well-matched to that of
WFPC2. Field centers were chosen to maximize the number of galaxies of
known redshift falling within the CISCO field of view. Given we are
concerned with counting satellites around individual hosts, our primary
results will not be biased by this maximization.

In total, six Groth Strip and four CFRS fields were imaged to a depth
($\sim$2.6 ks) deemed adequate for locating galaxies at least 2
magnitudes fainter than most of the hosts (see below). In total, 190
redshift survey galaxies were sampled in the \kband~(151 fully overlap
with HST images and are bright enough for the comparison to the optical
pair fraction). The infrared data were reduced using the AUTOMKIM
pipeline developed at the Subaru facility by the CISCO group.

The limiting depth for locating faint satellites was estimated by
performing photometry on artificial stars inserted into each image and
by comparing the observed galaxy number counts to those published by
\citet{djo95}.  Both techniques agree, demonstrating that the
CISCO data are complete at the 90\% level at $K=22.5$.  Object
detection and photometry in both the optical and infrared were carried
out using the SExtractor package \citep{ber96}.

\section{Optical versus Infrared-Selected Pair Fractions}

First, we compare the optical and infrared pair fractions, closely
following the precepts of \citet{lef00}, though we adopt the
cosmology---$\Omega_M = 0.3, \Lambda = 0.7$, and $h = 0.7$---instead of
$q_0 = 0.5$ and $h=0.5$ as used by \lefe~et al.  Pairs are identified as
companions to a limit no more than 1.5 magnitudes (independently in both
optical and \kband) fainter than their host galaxy within a separation
radius of $r_p=20$ kpc.  Satellites within this radius are expected to
strongly interact with the halo of the host and merge in less than
$\sim$1 Gyr due to dynamical friction \citep{pat97}.  Multiple
satellites around the same host are counted as separate pairs, and a
field correction is applied to the pair counts based on the observed
number density.  Throughout, we assume that the pair fraction is
independent of the intrinsic properties of the host galaxies and the way
they were selected.

The field-corrected optical and infrared pair fractions for a sample of
151 host galaxies of known redshift with $K<21$ and $I<23$ are presented
in Figure 1 and Table 1 and contrasted with the results using the same
optical procedure as derived from Table 3 of \citet{lef00}. Our first
redshift bin ($0.2<z<0.5$) contains too few hosts for useful
comparisons, but in the two higher redshift bins ($0.5<z<0.75$ and
$0.75<z<1.5$), the statistical significance is adequate. There, although
we find optical results comparable to \citet{lef00}, the infrared pair
fraction is a factor of 2-3 less.  See Figure 2 for some examples.

\begin{inlinefigure}
\begin{center}
\resizebox{\textwidth}{!}{\includegraphics{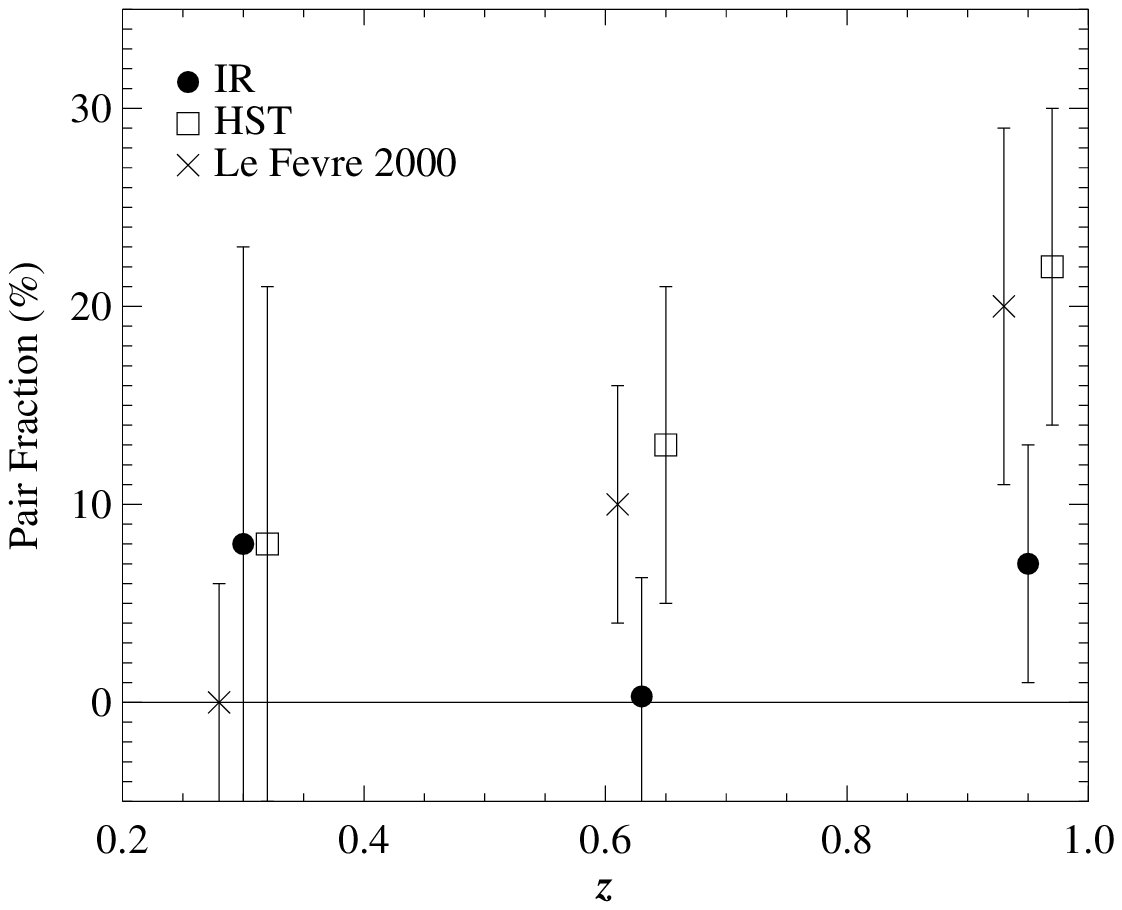}}
\end{center}
\figcaption{The field-subtracted pair fraction as measured in the
infrared and optical. \kband~measurements appear as filled circles
and \hst~F814W measures as squares. The new results are compared
with those of \citet{lef00}. \label{compare}}
\end{inlinefigure}

To examine the possibility that our comparison with an equivalent
\hst~analysis may be biased by resolution effects, we convolve each
$0\farcs1$ \hst~image to the corresponding CISCO resolution, which
varies from $0\farcs35$ to $0\farcs5$. We then repeat the detection and
analysis, including the background number counts. The \hst~pair fraction
decreases by only $\sim$30\%, remaining a factor of two above the
infrared pair fractions in the two highest redshift bins. We also
investigate the separation distribution between each optical companion
and its host. The smallest separation is just above $0\farcs5$, implying
that the majority of optically-identified pairs would be readily
resolved in the CISCO images, but were simply too faint in the infrared
to be counted. Both results suggest that resolution is not the primary
difference between the two samples, rather it is the bluer colors of the
satellite galaxies.  In general, observed satellite galaxies tend to be
bluer in $(V-K)$ than hosts, though the detection in IR favors redder
companions.

\begin{inlinefigure}
\begin{center}
\resizebox{\textwidth}{!}{\includegraphics{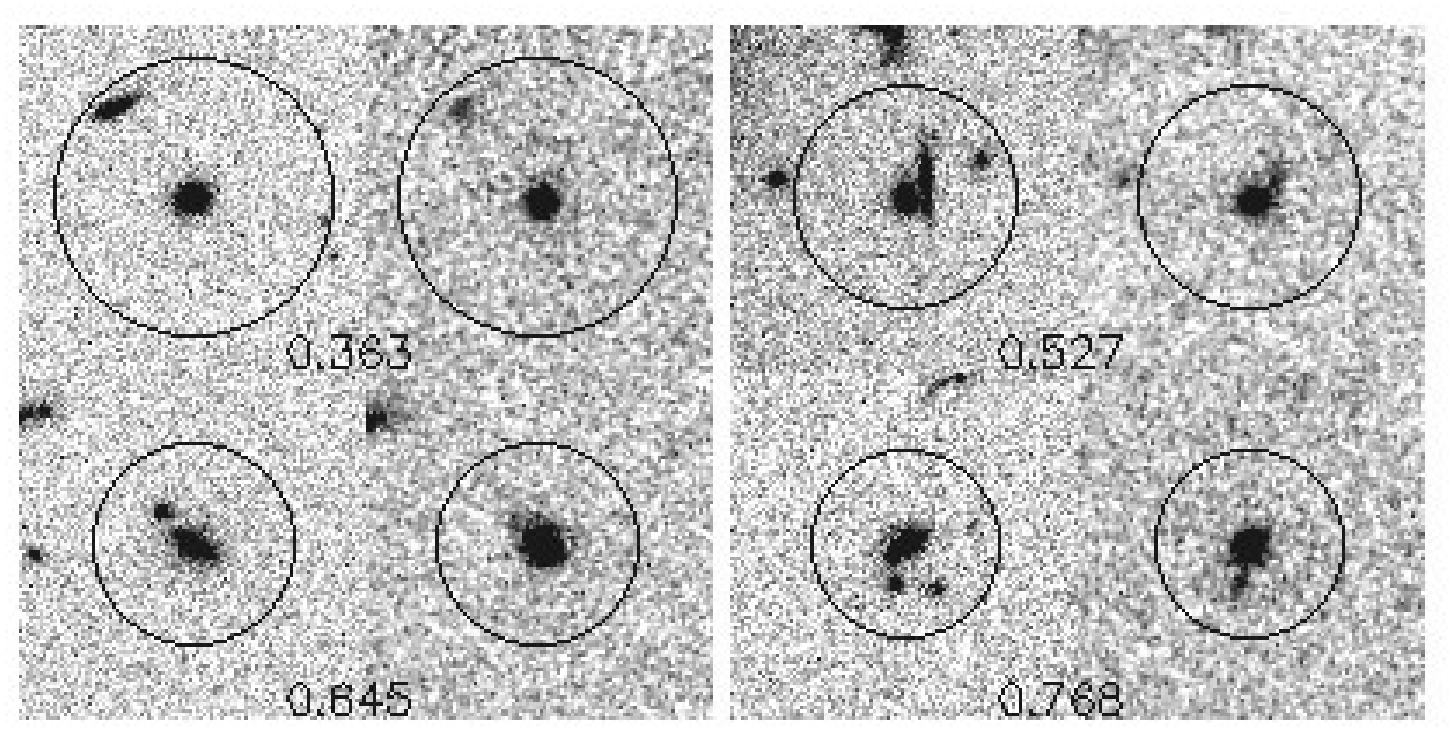}}
\end{center}
\figcaption{Examples of pairs identified in the optical but not in the
  infrared.  In each set, the left postage stamp is from \hst~$I_{814}$, and
  the right is from CISCO \kband~images.  The 20 kpc radius and
  redshifts are indicated. \label{ex}}
\end{inlinefigure}

\section{Weighted Infrared Pair Statistics}

We now explore a more detailed formalism for investigating the merger
history as laid out by \citet{pat00}.  Rather than applying a
differential magnitude cut linked to the host galaxy, we select
companions in a fixed absolute magnitude range, $-24 \leq M_{K'} \leq
-19$, regardless of the host.  This will increase the number of observed
pairs since we include fainter companion galaxies.  Our sample for this
analysis also grows to 190 hosts because host galaxies with $21.0 <
K'<22.5$ are now included.

As in $\S$3, the companion search radius is set to $r_p=20$ kpc and
field subtraction implemented as before. Absolute magnitudes are
calculated using $k$-corrections for the $K'$-band tabulated by
\citet{pog97} (using the Sa model), assuming that each companion galaxy
is at the same redshift as its host.

A volume-limited estimate of the pair fraction as a function of redshift
can be achieved by applying weights to both companions and hosts. As it
is easier to detect intrinsically fainter galaxies nearby, higher
redshift companions must be given more weight. The opposite is true for
host galaxies because the observed number of pairs per host is less
certain at higher redshift.  Following \citet{pat00}, these weights are
based on the comoving density of companions observed in a hypothetical
volume-limited survey compared to that in a flux-limited survey.  We
calculate these weights by integrating $K'$ (or $K_s$) band luminosity
functions from \citet{cole01}, for $z<0.6$, and \citet{kash03}, for
$0.6<z<1.5$.

The results are given in Table \ref{results2}.  At $z\sim 1$, the merger
fraction of 24\% $\pm$ 10\% is expectedly higher because of the
inclusion of fainter galaxies. The errors are derived from counting
statistics, and as in $\S$3, we expect the true pair fraction to be
slightly higher (about 1\%) than observed because some faint companions
may be obscured by large hosts.  The implied merger rate suggests that
35\% of typical $L^*$ galaxies have undergone a merger with a companion
in this luminosity range since $z \sim 1$.

\section{Mass Assembly Rates}

We have applied two pair counting methods and found that the Patton
\etal~method delivers a pair fraction higher than the technique of
\lefe~\etal~because the former includes fainter companions.  To
reconcile these two different results with a single mass assembly
history, we estimate the stellar mass accretion rate associated with
merging galaxies.  Because it is not known which companions are
physically associated with their host, the stellar mass of companions
can only be determined in a statistical sense.  We first fit our {\em
VIK$'$} photometry of host galaxies (with redshifts) to template SEDs
spanning a range of ages, star formation histories, and metallicities,
assuming a Salpeter IMF (with the range $0.1-100M_\odot$, Bruzual \&
Charlot 2000, private communication).  We then scale to the
\kband~luminosity to estimate the stellar mass (see Brinchmann et
al. 2000).  We assume the companions follow the same distribution of
$M_*/L_{K'}$ vs. $L_{K'}$ as the hosts and use this distribution to
estimate the stellar mass of companion galaxies.  Finally, though the
merging timescale depends on the details of the interaction, we follow
previous studies (e.g., Patton \etal~2000) and assume an average value
of 0.5 Gyr for galaxies separated by $r_p<20$ kpc.

\begin{inlinefigure}
\begin{center}
\resizebox{\textwidth}{!}{\includegraphics{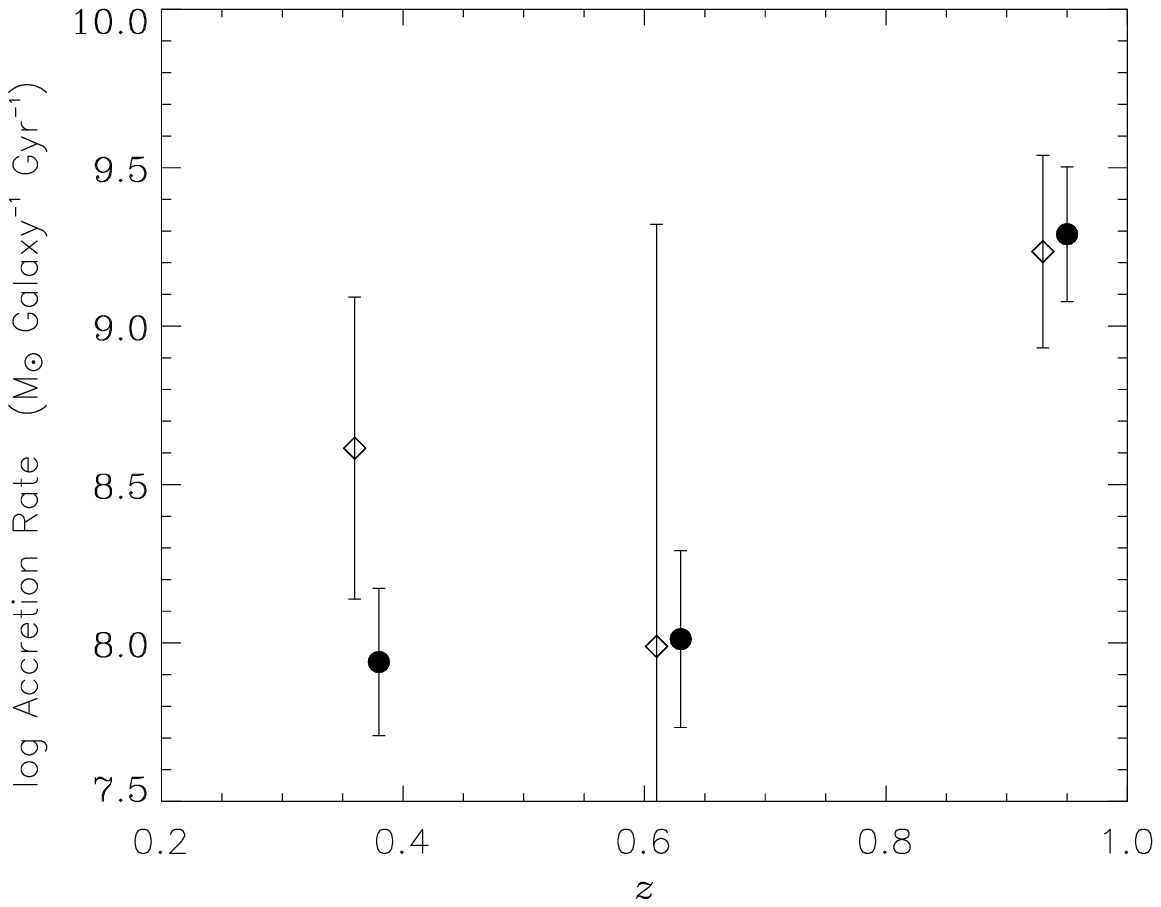}}
\end{center}
\figcaption{ Stellar mass accretion rate per galaxy in three redshift
  bins.  Filled symbols are the results from the Patton et al. weighted
  analysis.  Open symbols are the results of the \lefe~\etal~technique.
  The first redshift bin is again the least significant since it
  contains the fewest host galaxies. \label{macc}}
\end{inlinefigure}

With these assumptions, we demonstrate that the two very different pair
statistics are consistent with a similar merger history in terms of the
accreted stellar mass. In Figure \ref{macc}, open symbols are the mass
accretion rate from the \lefe~et al. method, and solid symbols are that
from the Patton et al. method.  The large error bars include both
statistics and 50\% uncertainties in $M_*/L_{K'}$. Both methods
illustrate a rise in the stellar mass accretion rate at the highest
redshifts, with the Patton et al. method giving a value of $2\times
10^{9 \pm 0.2}$ \msun~galaxy$^{-1}$ Gyr$^{-1}$ at $z \sim 1$. This mass
corresponds to $\approx$4\% of the average stellar mass of host galaxies
at these redshifts.  The result may be compared with an estimate made by
\citet{con03} of $6.4\times 10^{8 \pm 0.1}$ \msun~galaxy$^{-1}$
Gyr$^{-1}$ at $0.8<z<1.4$ using morphological indicators to distinguish
merger remnants.

We contrast our assembly rate from pre-existing stellar systems with the
{\em integrated} stellar mass density \citep{dick03}, which reflects the
growth of galaxies from newly-formed stars.  While not necessarily
completely independent (for example if merging triggers new star
formation), the relative magnitudes of the two phenomena are interesting
to consider.  Using the luminosity functions of Kashikawa et al (2003)
to determine the comoving number density of host galaxies, we integrate
the mass accretion rate to estimate the stellar mass assimilated by
galaxies in the host K-band luminosity range, finding $\Delta \rho_*^m
\approx 3 \times 10^{8 \pm 0.2}$ \msun~Mpc$^{-3}$.  For a Salpeter IMF,
we deduce that 30\% of the local stellar mass in luminous galaxies was
assimilated via merging of pre-existing stars since $z \sim 1$,
comparable to the build-up deduced by \citet{dick03} from ongoing star
formation.

\acknowledgments 

We thank Chris Simpson and Kentaro Aoki for their help during our
observations at the Subaru Telescope.  RSE and MF acknowledge the
generosity of the Japanese Society for the Promotion of Science.


%
%
\begin{deluxetable}{lcccccc}
\tablecaption{Pair Fraction}

\tablewidth{0pt}
\tablecolumns{7}
\tablehead{
\colhead{Sample} & \colhead{$z$} & \colhead{$N_{gal}$} &
\colhead{$N_{maj}$} &\colhead{$N_{proj}$} &\colhead{Pair Fraction (\%)}
&\colhead{Merger Fraction (\%)}}

\startdata
CISCO      &  0.2--0.5  &  22 & 6  (0.27) &  4.4 (0.20)  &  $8 \pm 14$  & $4 \pm 10$ \\
CISCO      &  0.5--0.75 &  47 & 4  (0.09) &  3.8 (0.08)  &  $0.3 \pm 6$
& $0.2 \pm 5$ \\
CISCO      &  0.75--1.5 &  74 & 11 (0.15) &  5.8 (0.08)  &  $7 \pm 6$
& $7 \pm 6$ \\
\hline
HST F814W  &  0.2--0.5  &  22  & 5  (0.23) & 3.3 (0.15)  &  $8 \pm 13$
& $5 \pm 8$ \\
HST F814W  &  0.5--0.75 &  47  & 10 (0.21) & 4.0 (0.09)  &  $13 \pm 8$
& $11 \pm 7$ \\
HST F814W  &  0.75--1.5 &  74  & 25 (0.34) & 8.5 (0.11)  &  $22 \pm 8$
& $21 \pm 8$ \\
\hline
\lefe & 0.2--0.5  & 98 & 11 (0.11) & 19   (0.19) & 0
& 0  \\
\lefe & 0.5--0.75 & 89 & 21 (0.24) & 12.2 (0.14) & $9.9 \pm 6$ & $8 \pm 5$ \\
\lefe & 0.75--1.3 & 62 & 21 (0.34) & 8.4  (0.14) & $20.3 \pm 9$ & $19.4
\pm 9$ \\
\enddata

\tablecomments{$N_{gal}$ is the total number of galaxies drawn from
the redshift sample.  $N_{maj}$ is the number of companions fitting
the pair criteria described in the text.  $N_{proj}$ is the expected
number of contaminating field galaxies.  The pair fraction is defined
as $(N_{maj}-N_{proj})/N_{gal}$ and the merger fraction is the
pair fraction corrected by a factor of $0.5(1+z)$ where $(1+z)$
corresponds to the mean redshift of the bin.  Numbers appearing in
parentheses are the averages per host galaxy, printed for comparison
to $N_c^D$ and $N_c^R$ in Table \ref{results2}.  Results from `All
CFRS+LDSS' in Table 3 of \citet{lef00} are also reproduced.  Errors are
calculated using counting statistics.\label{results1} }

\end{deluxetable}

%
%
\begin{deluxetable}{lcccccc}

\tablecaption{Weighted Pair Statistics}

\tablewidth{0pt}
\tablehead{
\colhead{$z$} & \colhead{$N_{gal}$} &
\colhead{$N_c^D$} & \colhead{$N_c^R$} & \colhead{$N_c$}
& \colhead{Average $M_K$} & \colhead{Merger Fraction}}

\startdata
0.2--0.5  &  30 & 0.18 & 0.04 & 0.14 $\pm$ 0.07 & -17.9 &
12 $\pm$ 5 \\
0.5--0.75 &  57 & 0.11 & 0.04 & 0.08 $\pm$ 0.06 & -18.3 &
8 $\pm$ 5 \\
0.75--1.5 &  93 & 0.29 & 0.03 & 0.26 $\pm$ 0.10 & -20.5 &
24 $\pm$ 10 \\
\enddata

\tablecomments{$N_{gal}$ is the total number of galaxies drawn from the
redshift sample.  $N_c^D$ is the raw, weighted number of companions per
host while $N_c^R$ is the projected fraction from the field.  The
corrected average is $N_c$.  The average $M_K$ is the associated \kband~luminosity in
companions averaged over every host in the sample.  The merger fraction
is calculated as before.  Errors are determined using weighted counting
statistics. \label{results2} }

\end{deluxetable}






\begin{thebibliography}{}
\bibitem[Baugh, Cole, \& Frenk(1996)]{baugh96} Baugh, G., Cole, S. \& Frenk, C. S. 1996, \mnras, 283, 1361 
\bibitem[Bertin \& Arnouts(1996)]{ber96} Bertin, E.\& Arnouts, S. 1996, \aap, 117,393 
\bibitem[Brinchmann \& Ellis(2000)]{brinch00} Brinchmann, J. \& Ellis,
  R. S. 2000, \apj, 546, L77 
\bibitem[Brinchmann et al.(1998)]{brinch98} Brinchmann, J., et al. 1998, \apj, 499, 112 
\bibitem[Broadhurst, Ellis, \& Glazebrook(1992)]{broad92} Broadhurst, T. J., Ellis,
  R. S., \& Glazebrook, K. 1992, \nat, 355, 55
\bibitem[Burkey, Keel, \& Windhorst(1994)]{burk94} Burkey, J. M., Keel, W. C.,
  \& Windhorst, R. A. 1994, \apj, 429, 13
\bibitem[Carlberg, Prichet, \& Infante(1994)]{carl94} Carlberg, R. G., Pritchet,
  C. J., \& Infante, L. 1994, \apj, 435, 540
\bibitem[Cole et al.(2001)]{cole01} Cole, S., et al. 2001, \mnras, 326,
  255 
\bibitem[Conselice et al.(2003)]{con03} Conselice, C. J., Bershady,
  M. A., Dickinson, M., \& Papovich, C. 2003, \apj, 126, 1183
\bibitem[Dickinson et al.(2003)]{dick03} Dickinson, M., Papovich, C.,
  Ferguson, H. C., \& Bud\'{a}vari, T. 2003, \apj, 587, 25 
\bibitem[Djorgovski et al.(1995)]{djo95} Djorgovski, S. et al. 1995, \apjl, 438, L13 
\bibitem[Ellis et al.(1996)]{ellis96} Ellis, R. S., Colless, M.,
  Broadhurst, T., Heyl, J., \& Glazebrook, K. 1996, \mnras, 280, 235
\bibitem[Giavalisco et al.(1996)]{gia96} Giavalisco, M., Livio, M.,
  Bohlin, R. C., Macchetto, F. D., \& Stecher, T. P. 1996, \aj, 112, 369
\bibitem[Groth et al.(1994)]{gro94} Groth, E. J., Kristian, J. A.,
  Lynds, R., O'Neil, E. J., Balsano, R., \& Rhodes, J. 1994, \baas, 26, 1403
\bibitem[Kashikawa et al.(2003)]{kash03} Kashikawa, N., et al. 2003, \aj, 125, 53 
\bibitem[Kauffmann(1996)]{kauf96} Kauffmann, G. 1996, \mnras, 281, 475
\bibitem[Koo(1995)]{koo95} Koo, D. C. 1995, in Wide Field Spectroscopy
  and the Distant Universe, ed. S. Maddox \& Arag\'{o}n-Salamanca
  (Singapore: World Scientific), 55,
\bibitem[\lefe~et al.(2000)]{lef00} \lefe, O., et al. 2000, \mnras, 311, 565 
\bibitem[Lilly et al.(1995)]{lilly95} Lilly, S. J., Le Fevre, O.,
  Crampton. D., Hammer, \& F., Tresse, L. 1995, \apj, 455, 50
\bibitem[Lilly et al.(1996)]{lilly96} Lilly, S. J., Le Fevre, O.,
  Hammer, \& F., \& Crampton. D., 1996, \apjl, 460, L1
\bibitem[Motohara et al.(2002)]{moto02} Motohara, K., et al. 2002, \pasj, 54, 315 
\bibitem[Patton et al.(2000)]{pat00} Patton, D. R., Carlberg, R. G.,
  Marzke, R. O., Pritchet, C. J., da Costa, L. N., \& Pellegrini, P. S. 2000, \apj, 536, 153 
\bibitem[Patton et al.(2002)]{pat02} Patton, D. R., et al. 2002, \apj, 565, 208 
\bibitem[Patton et al.(1997)]{pat97} Patton, D. R., Pritchet, C. L.,
  Yee, H. K. C., Ellingson, E., \& Carlberg, R. G. 1997, \apj, 475, 29 
\bibitem[Poggianti(1997)]{pog97} Poggianti, B. M. 1997, \aap, 122, 399 
\bibitem[Yee \& Ellingson(1995)]{yee95} Yee, H. K. C., \& Ellingson,
  E. 1995, \apj, 445, 37
\end{thebibliography}
\end{document}